# TRACKING A DECADE OF RESEARCH AT THE UNIVERSITY OF NIGERIA, NSUKKA: A SCIENTOMETRIC ANALYSIS (2014-2023)

Muneer Ahmad[1] and Joseph U. Igligli[2]

[1]Kampala International University, Uganda

[2]University of Calabar, Nigeria

muneerbangroo@gmail.com

**Abstract**

*This study employs scientometric methods to assess the research output and performance of the University of Nigeria from 2014 to 2023. By analyzing publication trends, citation patterns, and collaboration networks, the research aims to comprehensively evaluate the university's research productivity, impact, and disciplinary focus. These research endeavors are characterized by innovation, interdisciplinary collaboration, and commitment to excellence, making the University of Nigeria a significant hub for cutting-edge research in Nigeria and beyond. The present study has been undertaken to determine the impact of the university's research and publication trends from 2014 to 2023. The study focuses on year-wise research output, citation impact at local and global levels, prominent authors and their total output, top journals, collaborating countries, and the most contributing departments of the University of Nigeria. The university's ten years of publication data indicate that 6,353 papers were published from 2014 to 2023, receiving 86,202 citations with an h-index of 39. In addition to this, the stenographical mapping of data is presented through graphs using the VOSviewer software mapping technique. The findings of this study will contribute to understanding the university's research strengths, weaknesses, and potential areas for improvement. Additionally, the results will inform evidence-based decision-making for enhancing research strategies and policies at the University of Nigeria.*

## INTRODUCTION

The University of Nigeria, commonly called UNN, is a premier academic institution in Nsukka, Enugu State. Established in 1960 by Dr. Nnamdi Azikiwe, it holds the distinction of being Nigeria's first fully autonomous university. The founding vision was to emulate the American educational system, providing a broad-based education that includes not only academic excellence but also practical skills and ethical grounding (University of Nigeria, n.d.).

UNN operates multiple campuses, including the Nsukka campus, Enugu campus, and the Ituku-Ozalla campus, each dedicated to various faculties and specialties. The university offers various undergraduate and postgraduate programs across its 15 faculties, including Arts, Biological Sciences, Education, Engineering, and Social Sciences (Okoro, 2013). This diversity in academic offerings underscores UNN's commitment to providing comprehensive education tailored to the needs of Nigeria's diverse economy and society.

A notable feature of the University of Nigeria is its commitment to research and innovation. The institution houses several research centers and institutes, such as the Institute for Development Studies, the Centre for Entrepreneurship and Development Research, and the Roar Nigeria Hub, a technology and innovation hub that fosters tech startups (Eze, 2019). These centers enhance the university's research capabilities and contribute significantly to national development by addressing critical issues and generating practical solutions.

The University of Nigeria has a strong alumni network and has significantly contributed to various fields within Nigeria and internationally. Notable alumni include Chinua Achebe, a celebrated novelist and author of "Things Fall Apart," and Charles Soludo, a former Governor of the Central Bank of Nigeria (Ogbonna, 2020). The university's emphasis on leadership and service is reflected in the achievements of its graduates, who continue to play pivotal roles in academia, governance, business, and the arts.

Furthermore, UNN places a high priority on international collaborations and partnerships. It has established exchange programs and collaborative research initiatives with numerous universities and research institutions worldwide. This global engagement enhances the quality of education and research at UNN and provides students and faculty with opportunities for cross-cultural learning and exposure to international best practices (Agu, 2018).

Despite its many achievements, the University of Nigeria faces challenges common to many African higher education institutions, including inadequate funding, infrastructural deficits, and brain drain. However, the university strives for excellence by seeking innovative solutions and leveraging its strong alumni network and international partnerships to overcome these hurdles (Nweze, 2021). Its resilience and ongoing efforts to improve facilities, curriculum, and research output highlight its unwavering dedication to its founding principles and future growth.

## REVIEW OF LITERATURE

Many researchers employ bibliometric methods using international databases to extract and analyze research data for diverse purposes. Some focus on country or institutional performance. Darmadji et al. (2018) analyzed the Islamic University of Indonesia’s publications in Scopus (2005-2017) to assess author productivity, collaboration, and comparisons with other universities. Findings revealed steady growth in output but limited collaboration, recommending incentives and wider international engagement. Similarly, Ahmad and Batcha (2019) studied Bharathiar University’s publications (2009–2018), reporting 3,440 papers with 38,104 citations and an h-index of 68. Using VOSviewer, they identified major contributors, top journals, and international collaborators.

Mahala and Singh (2021) investigated the research output of top Indian universities (2015–2019) in the Web of Science. Among Delhi, Banaras Hindu, Anna, Jadavpur, and Punjab Universities, Delhi produced the highest output. Results showed multi-authored papers attracted more citations, with strong collaborations involving the USA, South Korea, and Germany. Ahmad (2022) examined coronary artery disease research among BRICS countries (1990–2019), showing continuous growth, dominant multi-authorship, and China’s leading productivity, with English as the main communication language.

(Matthews, 2013) assessed South African physicists' productivity (2009–2011) using Web of Science data. Median productivity was 1.33 papers annually, with professors outperforming lecturers. South African output was comparable to mid-ranked U.S. institutions but below top-ranked global universities. Kumar and Dora (2012) studied IIMA's publications, focusing on publication types, preferred journals, and authorship patterns, while Baskaran (2013) analyzed Alagappa University's output, identifying author productivity and institutional collaborations.

Das et al. (2021) explored Mizoram University's publications (2002–2018), totaling 586 papers—mostly research articles - with Thapa, R.K., and Tiwari, D. as leading contributors. The most active years were 2016 and 2017, and Current Science was the top publishing venue. Recent studies on African universities include Ahmad and Nkatv (2025), who analyzed the University of Ibadan's output (2014–2023), reporting 7,159 publications, 218,572 citations, and an h-index of 75, revealing collaboration gaps through VOSviewer visualization. Likewise, Ahmad and Ubi (2025) examined the University of Lagos (2004–2023), noting consistent growth, particularly in Health Sciences, Engineering, and Social Sciences, with strong U.S. and U.K. collaborations enhancing global visibility.

**Objectives of the Study**

1. To look at the research output growth pattern of the University of Nigeria, Nsukka, from 2014 to 2023.
2. Discover which University of Nigeria, Nsukka authors are the most prolific.
3. Determine which esteemed journals publish academic works from the University of Nigeria.
4. To identify and rank the major institutions collaborating with the University of Nigeria, Nsukka, from 2014 to 2023, highlighting patterns and strengths of research partnerships.
5. To analyze and rank departmental research productivity within the University of Nigeria, Nsukka, over the decade, revealing disciplinary contributions and distribution of scholarly output.
6. To examine the distribution of UNN publications by document type and assess how different formats contributed to overall scholarly communication during the study period.
7. To identify and rank the most relevant countries of corresponding authors associated with the University of Nigeria, Nsukka publications, mapping the geographic breadth and international engagement of the university's research.

**METHODOLOGY**

The current study uses the Clarivate Analytics Web of Science database to quantify the University of Nigeria's research productivity. Using a basic search, the information was obtained from the Web of Science (WOS) database (https://www.webofscience.com/wos/woscc/basic-search). By searching for the term " University of Nigeria " in the affiliation field with the date span set to 2014–2023 and the indexes SCI–EXPANDED, SSCI, and AHCI, publications published by the University of Nigeria were found on WOS. A total of 6,353 articles were located, and Histcite, VOS viewer, and MS-Excel software programs were used to examine the gathered data. Bibliometric tools and techniques have carefully examined the computed data to obtain the intended outcome that satisfies the study objectives.

**RESULTS AND DISCUSSION**

**Data for the Study**

**Table 1:** Main Information about Data

| S/N | Description | Results |
|---|---|---|
| 1 | Timespan | 2014-2023 |
| 2 | Sources (Journals, Books, etc.) | 1,934 |
| 3 | Documents | 6,353 |
| 4 | Annual Growth Rate % | -8.6 |
| 5 | Document Average Age | 4.22 |
| 6 | Average citations per doc | 13.57 |
| 7 | References | 242,347 |
| | DOCUMENT CONTENTS | |
| 8 | Keywords Plus (ID) | 11,097 |
| 9 | Author's Keywords (DE) | 16,341 |
| | AUTHORS | |
| 10 | Authors | 35,659 |
| 11 | Authors of single-authored docs | 204 |
| | AUTHORS COLLABORATION | |
| 12 | Single-authored docs | 361 |
| 13 | Co-Authors per Doc | 12.5 |
| 14 | International co-authorships % | 46.88 |

As presented in Table 1, a total of 35,659 authors contributed 6,353 documents with a total of 242,347 references, research from 1934 Scopus-indexed journals, including articles, conference papers, editorials, and review papers published from 2014 to 2023, which spread around nine years. The average number of co-authors per article was 12.5, and only 204 articles were published by single authors, indicating that multi-authorship is common in research.

**Evaluation of the Annual Output of Publications of the University of Nigeria**

**Table 2:** Research Output Growth Pattern of the University of Nigeria, Nsukka (2014–2023)

| S/N | Year | Records | % | Rank | TLCS | % | TGCS | % |
|---|---|---|---|---|---|---|---|---|
| 1 | 2014 | 312 | 4.91 | 9 | 332 | 7.92 | 4,035 | 4.68 |
| 2 | 2015 | 292 | 4.60 | 10 | 311 | 7.42 | 3,645 | 4.23 |
| 3 | 2016 | 414 | 6.52 | 8 | 386 | 9.21 | 6,221 | 7.22 |
| 4 | 2017 | 453 | 7.13 | 7 | 434 | 10.35 | 13,790 | 16.00 |
| 5 | 2018 | 454 | 7.15 | 6 | 448 | 10.68 | 6,034 | 7.00 |
| 6 | 2019 | 614 | 9.66 | 5 | 651 | 15.53 | 13,005 | 15.09 |
| 7 | 2020 | 789 | 12.42 | 4 | 739 | 17.62 | 19,620 | 22.76 |
| 8 | 2021 | 863 | 13.58 | 3 | 441 | 10.52 | 8,823 | 10.24 |
| 9 | 2022 | 984 | 15.49 | 2 | 267 | 6.37 | 7,404 | 8.59 |
| 10 | 2023 | 1,178 | 18.54 | 1 | 184 | 4.39 | 3,625 | 4.21 |
| Total | | **6,353** | 100.00 | | **4,193** | 100.00 | **86,202** | 100.00 |

TLCS = Total Local Citation Score; TGCS = Total Global Citation Score

Table 2 presents the number of articles, total citations, and average citations per year for various authors' articles per year; the data includes the number of articles published, the number of citations they have received, and the annual average number of citations. We found that 6,353 research articles were published from 2014 to 2023 (Web of Science database). The highest number of articles was published in 2023 was 1,178 (18.54%), whereas the least was published in 2015, with 292 (4.60%). There are variations in the number of articles published during the period. According to the year-wise analysis, we found that 2016, 2017, 2018, and 2020 had more than 400 publications. We examined the Total Local Citation Score (TLCS) and noted that 2020 had the highest TLCS with 19,620. The second-highest number of TLCS, 13,790, was in 2017. The remaining years had TLCS values below 13,005. The researchers have also examined the Total Global Citation Scores (TGCS) values of overall periods of 86,202 citation scores. In 2020, the TGCS value received the highest number of 739; in 2023, it was the lowest at 184.

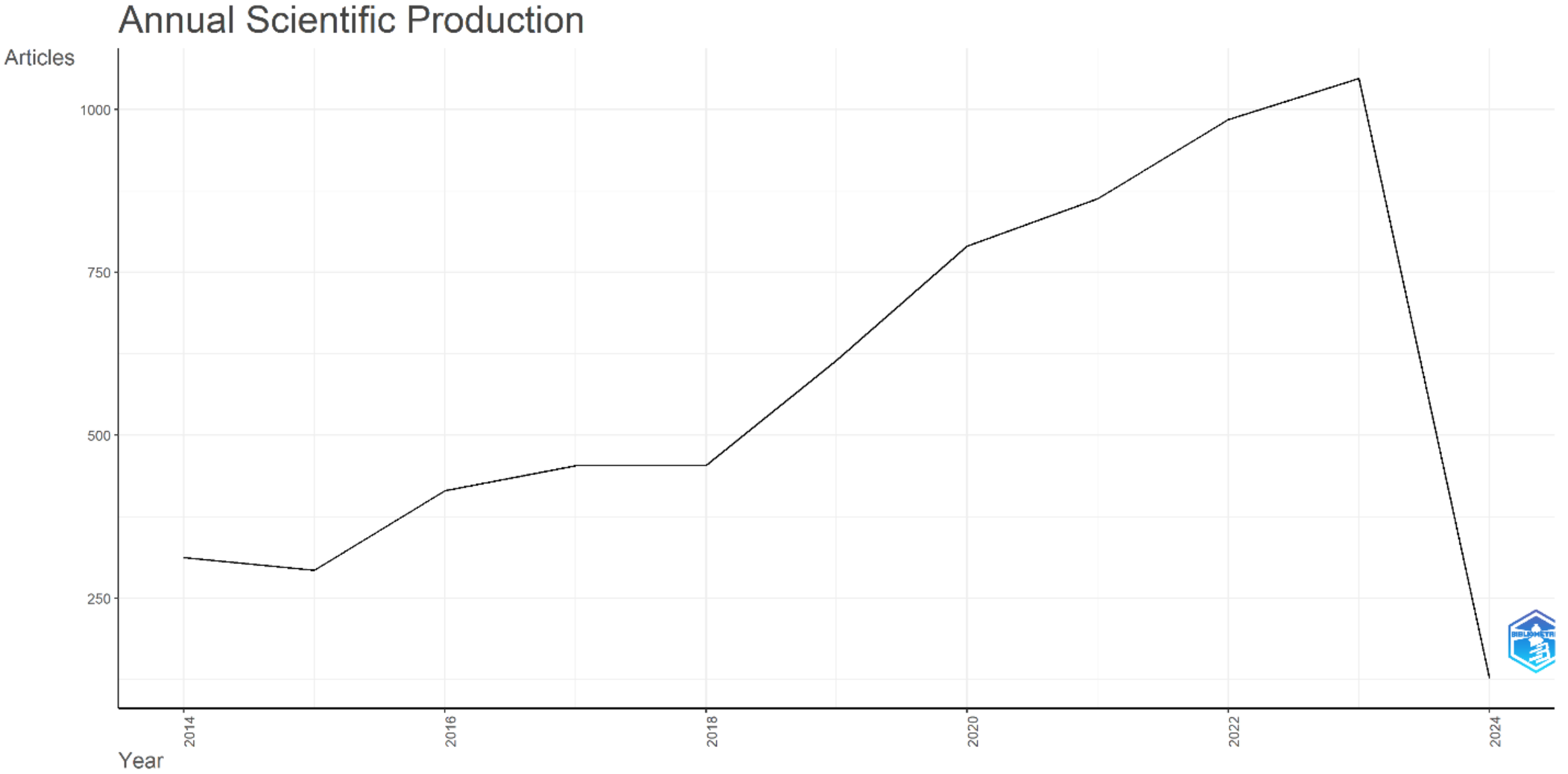

Figure 1: Annual Scientific Production

### Analysis of the Publication Output of the Top 20 Authors of the University of Nigeria

**Table 3:** Most Prolific Authors of the University of Nigeria, Nsukka and their Citation Score

| S/N | h-index | Authors | Citation sum within h-core | All citations | All articles |
|---|---|---|---|---|---|
| 1 | 34 | Mokdad, A. H. | 23,428 | 23,503 | 39 |
| 2 | 33 | Fischer, F. | 23,347 | 23,405 | 37 |
| 3 | 33 | Ezema, F. I. | 2,104 | 3,371 | 144 |
| 4 | 33 | Yonemoto N. | 23,347 | 23,431 | 39 |
| 5 | 33 | Onwujekwe, O.E. | 17,304 | 17,677 | 72 |
| 6 | 32 | Shaikh, M. A. | 23,105 | 23,189 | 38 |
| 7 | 31 | Hay, S. I. | 18,783 | 18,838 | 35 |
| 8 | 30 | Kisa, A. | 17,028 | 17,103 | 35 |
| 9 | 30 | Samy, A. M. | 19,214 | 19,298 | 36 |
| 10 | 30 | Mohammed, S. | 21,269 | 21,395 | 40 |
| 11 | 30 | Maaza, M. | 1,932 | 2,764 | 95 |

| 12 | 29 | Singh, J. A. | 19,312 | 19,396 | 35 |
|---|---|---|---|---|---|
| 13 | 29 | Murray, C. J. L. | 21,944 | 21,980 | 31 |
| 14 | 29 | Arabloo, J. | 16,986 | 17,070 | 35 |
| 15 | 29 | Jonas, J. B. | 23,016 | 23,062 | 33 |
| 16 | 29 | Koyanagi, A. | 19,509 | 19,584 | 34 |
| 17 | 29 | Haj-Mirzaian A. | 27,346 | 27,405 | 33 |
| 18 | 28 | Islam,, S. M. S. | 19,014 | 19,068 | 33 |
| 19 | 28 | Aljunid, S. M. | 16,029 | 16,054 | 30 |
| 20 | 28 | Younis, M. Z. | 18,950 | 18,992 | 31 |

Table 3 measures the impact of various authors by looking at several different metrics. It presents the top 20 authors and citation scores based on the number of published articles and their h-index. It also displays the top 20 most cited publications. The data shows the number of articles published by each author. The data shows that the most relevant author is Mokdad AH, with 23,428 citations, a h-core of 23,503 articles published, and an h-index of 39; this is followed by Fischer F, with 23,347 citations, an h-score of 23,405 articles published, and an h-index of 37, and Yonemoto N, with 23,347 Citations sum with h-core, and an h-index 23,431 and 39 articles published. The other authors on the list include Shaikh MA, Jonas JB, and Mohammed S. This data indicates that these authors are leading researchers and have published many articles.

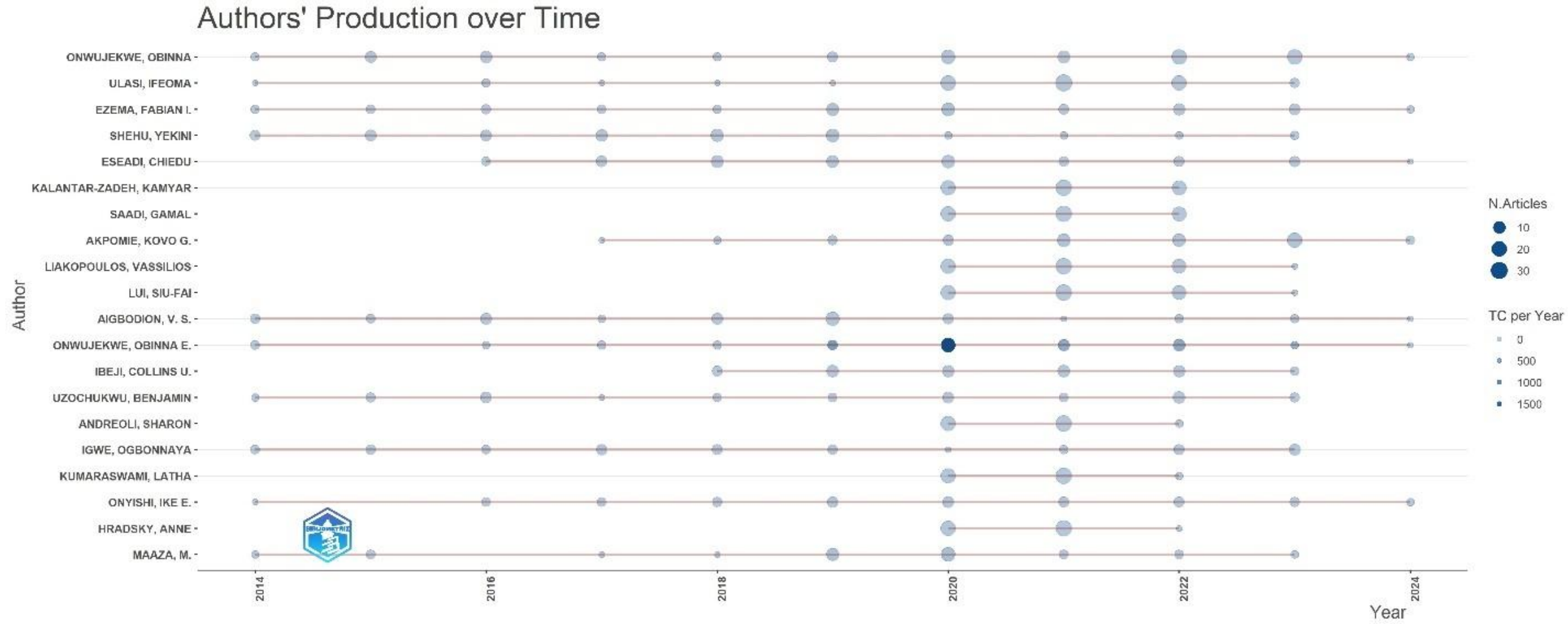

Figure 2: Authors' Production

It measures the impact of various authors by looking at several different metrics. The h-index measures the productivity and citation impact of a scientist or scholar. The g-index measures a scientist's productivity and impact, considering the number of highly cited papers they have. The TC (total citations) and NP (number of publications) are also included as metrics of an author's impact. The table provided lists the authors in descending order of h-index. However, it is essential to note that different metrics can give different perspectives on an author's impact, and it is usually recommended to consider multiple metrics to get a more complete picture.

### Analysis of Source-Wise Distribution of Documents

**Table 4:** Leading Journals Publishing Research from the University of Nigeria, Nsukka

| S/N | Sources | Documents | h_index | g_index | m_index | TC | PY Start |
|---|---|---|---|---|---|---|---|
| 1 | Lancet | 21 | 26 | 1.909 | 14,989 | 26 | 2014 |
| 2 | Medicine | 16 | 22 | 1.778 | 812 | 100 | 2016 |
| 3 | Applied Water Science | 15 | 26 | 2.143 | 740 | 26 | 2018 |
| 4 | Nigerian Journal of Clinical Practice | 15 | 23 | 1.364 | 1,304 | 264 | 2014 |
| 5 | PLOS One | 15 | 21 | 1.364 | 654 | 74 | 2014 |
| 6 | Environmental Science and Pollution Research | 14 | 33 | 1.75 | 1,091 | 35 | 2017 |
| 7 | Heliyon | 14 | 22 | 2.333 | 716 | 74 | 2019 |
| 8 | Environmental Monitoring and Assessment | 13 | 24 | 1.182 | 693 | 50 | 2014 |
| 9 | Scientific Reports | 13 | 23 | 2.167 | 585 | 45 | 2019 |
| 10 | BMC Public Health | 12 | 17 | 1.091 | 333 | 31 | 2014 |
| 11 | Journal Of Rational-Emotive and Cognitive-Behavior Therapy | 12 | 17 | 1.333 | 347 | 34 | 2016 |
| 12 | Molecules | 12 | 21 | 1.091 | 649 | 21 | 2014 |
| 13 | African Health Sciences | 11 | 18 | 1 | 611 | 104 | 2014 |
| 14 | Journal of Environmental Management | 11 | 13 | 1.222 | 718 | 13 | 2016 |
| 15 | BMC Health Services Research | 10 | 18 | 0.909 | 348 | 30 | 2014 |
| 16 | International Journal of Energy Research | 10 | 15 | 2 | 252 | 16 | 2020 |
| 17 | Journal of African Earth Sciences | 10 | 18 | 0.909 | 368 | 35 | 2014 |
| 18 | Journal of Molecular Structure | 10 | 17 | 1.111 | 346 | 34 | 2016 |
| 19 | Renewable & Sustainable Energy Reviews | 10 | 12 | 1 | 405 | 12 | 2015 |
| 20 | Arabian Journal of Chemistry | 9 | 13 | 0.818 | 202 | 13 | 2014 |

TC = Total Citations

Table 4 measures the impact of sources by looking at several different metrics. The h-index measures the productivity and citation impact of a scientist or scholar. The g-index measures a scientist's productivity and impact, taking into account the number of highly cited papers. The TC (total citations) and NP (number of publications) are also included as metrics of an author's impact. The table provided lists the authors in descending order of h-index, with The Lancet having the highest h-index, 26, followed by Medicine with an h-index of 22, and the Arabian Journal of Chemistry having the lowest (13). However, when looking at other metrics, such as the g-index, the Lancet has the highest g-index (21), followed by Medicine (16), and the Arabian Journal of Chemistry has the lowest (9). It is important to note that different metrics can provide different perspectives on an author's impact, and it is usually recommended to consider multiple metrics to get a more complete picture.

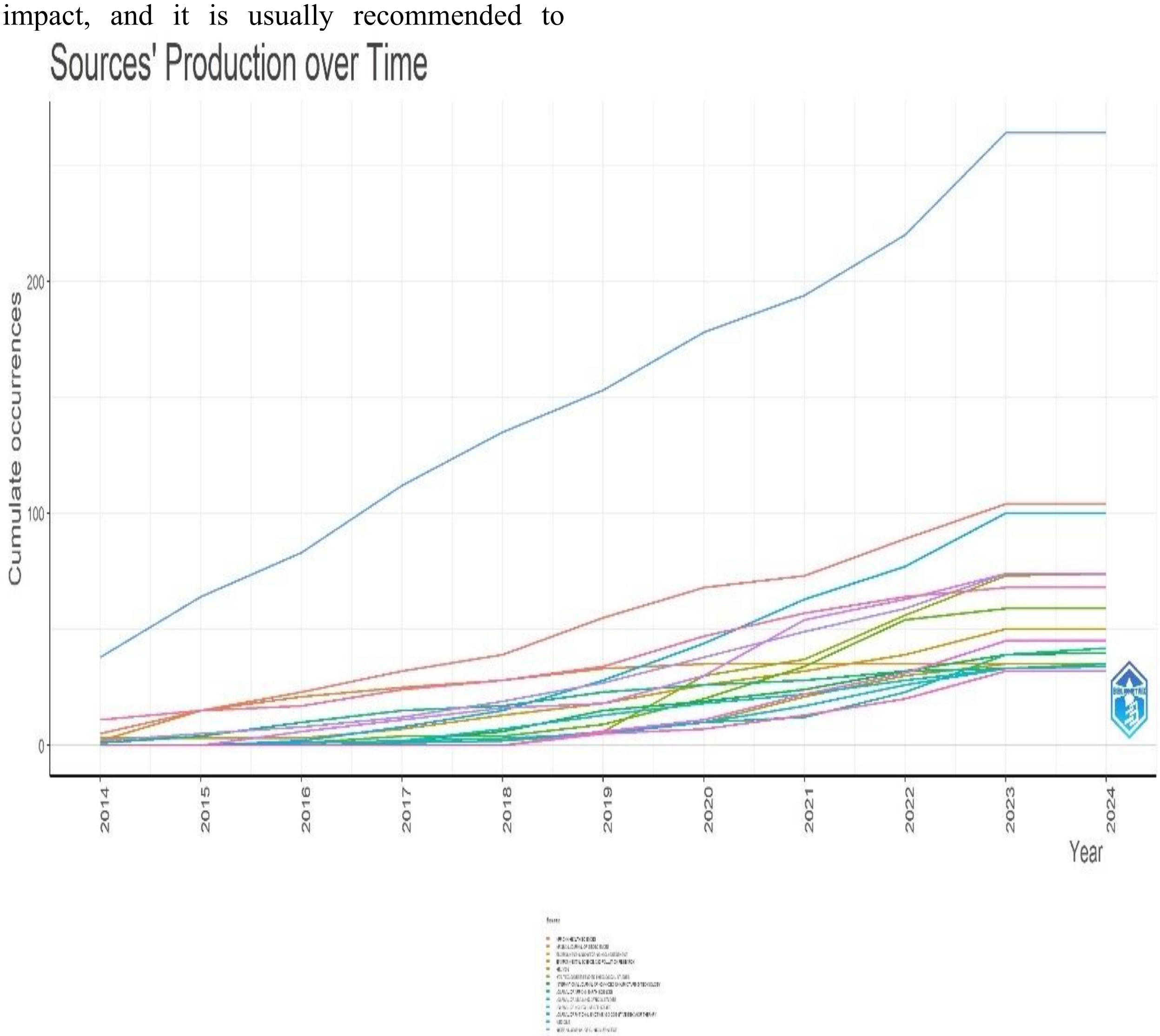

Figure 3: Source-wise Production

### Analysis of Collaborative Ranking of Institutions

**Table 5:** Major Institutional Collaborations of the University of Nigeria, Nsukka (2014–2023)

| S/N | Institution | Records | % | TLCS | TGCS | ACPP |
|---|---|---|---|---|---|---|
| 1 | University of Nigeria, Nsukka | 426 | 6.71 | 298 | 23,419 | 54.97 |
| 2 | Nnamdi Azikiwe University | 385 | 6.06 | 200 | 4,101 | 10.65 |
| 3 | University of Ibadan | 199 | 3.13 | 133 | 18,489 | 92.91 |
| 4 | University of Nigeria Teaching Hospital | 190 | 2.99 | 41 | 1,895 | 9.97 |
| 5 | University of Lagos | 180 | 2.83 | 148 | 22,831 | 126.84 |

| 6 | University of Johannesburg | 175 | 2.75 | 154 | 1,631 | 9.32 |
|---|---|---|---|---|---|---|
| 7 | Ebonyi State University | 150 | 2.36 | 117 | 1,288 | 8.59 |
| 8 | Alex Ekwueme Federal University, Ndufu-Alike | 148 | 2.33 | 128 | 1,564 | 10.57 |
| 9 | Ahmadu Bello University | 142 | 2.24 | 144 | 22,337 | 157.30 |
| 10 | University of the Free State | 142 | 2.24 | 271 | 10,630 | 74.86 |
| 11 | Enugu State University of Science and Technology | 139 | 2.19 | 79 | 1,047 | 7.53 |
| 12 | University of Calabar | 130 | 2.05 | 70 | 1,800 | 13.85 |
| 13 | University of KwaZulu-Natal | 123 | 1.94 | 185 | 23,992 | 195.06 |
| 14 | Jiangsu University | 122 | 1.92 | 62 | 2,430 | 19.92 |
| 15 | University of Benin | 122 | 1.92 | 101 | 11,838 | 97.03 |
| 16 | Federal Medical Centre | 120 | 1.89 | 52 | 2,521 | 21.01 |
| 17 | London School of Hygiene & Tropical Medicine | 119 | 1.87 | 136 | 20,572 | 172.87 |
| 18 | Cairo University | 117 | 1.84 | 113 | 23,662 | 202.24 |
| 19 | The Chinese University of Hong Kong | 115 | 1.81 | 106 | 13,167 | 114.50 |
| 20 | Michael Okpara University of Agriculture, Umudike | 110 | 1.73 | 88 | 1,097 | 9.97 |

TLCS = Total Local Citation Score; TGCS = Total Global Citation Score; ACPP= Average Citations Per Paper

Table 5 shows the ranking of collaborative institutions with the University of Nigeria. Within the period of study, contributors from 20 institutions have a total of 3,232 research collaborations with University of Nigeria among these institutions, the highest collaboration is recorded by the University of Nigeria Nsukka 426 (6.71%) followed by Nnamdi Azikiwe University 385 (6.06%), University Ibadan 199 (3.13 University Nigeria Teaching Hospital 190 (2.99%), University Lagos 180 (2.83%), Govt University Johannesburg 175 (2.75%), Ebonyi State University 150 (2.36%), Alex Ekwueme Fed University 148 (2.33%), Ahmadu Bello University 142 (2.24%), the least scored is by Michael Okpara University Agr 110 (1.73%). The analysis shows that the University of Nigeria, Nsukka, is the highest collaborating University with the University of Nigeria.

## Analysis of the Ranking of Department-wise Distribution

**Table 6:** Departmental Research Productivity at the University of Nigeria, Nsukka

| S/N | Institution with Sub-Division | Records | % | TLCS | TGCS | ACPP |
|---|---|---|---|---|---|---|
| 1 | University of Nigeria, College of Medicine | 538 | 8.5 | 234 | 4,778 | 8.88 |
| 2 | University of Nigeria Teaching Hospital | 388 | 6.1 | 162 | 4,384 | 11.30 |
| 3 | University of Nigeria, Department of Pure and Industrial Chemistry | 302 | 4.8 | 497 | 4,593 | 15.21 |
| 4 | University of Nigeria, Department of Physics and Astronomy | 284 | 4.5 | 356 | 4,242 | 14.94 |
| 5 | University of Nigeria | 237 | 3.7 | 60 | 1,548 | 6.53 |

| 6 | University of Nigeria, Department of Psychology | 199 | 3.1 | 199 | 2,104 | 10.57 |
|---|---|---|---|---|---|---|
| 7 | University of Nigeria, Department of Geology | 177 | 2.8 | 203 | 1,679 | 9.49 |
| 8 | University of Nigeria, Faculty of Physical Sciences | 161 | 2.5 | 157 | 2,067 | 12.84 |
| 9 | University of Nigeria, Faculty of Biological Sciences | 159 | 2.5 | 87 | 2,344 | 14.74 |
| 10 | University of Nigeria, Faculty of Pharmaceutical Sciences | 148 | 2.3 | 51 | 1,508 | 10.19 |
| 11 | University of Nigeria, Department of Educational Foundations | 128 | 2 | 204 | 894 | 6.98 |
| 12 | University of Nigeria, Department of Microbiology | 128 | 2 | 53 | 1,972 | 15.41 |
| 13 | University of Nigeria, Department of Metallurgical and Materials Engineering | 117 | 1.8 | 170 | 1,656 | 14.15 |
| 14 | University of Nigeria, Department of Zoology and Environmental Biology | 111 | 1.8 | 96 | 856 | 7.71 |
| 15 | University of Nigeria, Department of Mathematics | 110 | 1.7 | 110 | 1,751 | 15.92 |
| 16 | University of Nigeria, Department of Mechanical Engineering | 110 | 1.7 | 147 | 1,606 | 14.60 |
| 17 | University of Nigeria, Faculty of Social Sciences | 100 | 1.6 | 46 | 367 | 3.67 |
| 18 | University of Nigeria, Department of Electrical Engineering | 96 | 1.5 | 40 | 733 | 7.64 |
| 19 | University of Nigeria, Faculty of Veterinary Medicine | 94 | 1.5 | 40 | 466 | 4.96 |
| 20 | University of Nigeria, School of General Studies | 94 | 1.5 | 17 | 802 | 8.53 |

TLCS = Total Local Citation Score; TGCS = Total Global Citation Score; ACPP= Average Citations Per Paper

Table 6 shows the ranking-wise distribution of 20 departments from the University of Nigeria. From the analysis, it is discovered that the University of Nigeria, College of Medicine records the highest ranking of Department distribution with (538), while the University of Nigeria, Teaching Hospital came second with (338), and the University of Nigeria, Department of Pure & Ind Chem (302). This reveals the ranking Department-wise distribution of the University of Nigeria, College of Medicine as the highest at (8.5%), whereas the University of Nigeria, Teaching Hospital recorded (at 6.1%), the University of Nigeria, Department of Pure & Ind Chem (at 4.8%). Other universities in Nigeria, such as the Department Phys & Astron, the University of Nigeria, and the University of Nigeria, Department Psychol, recorded significant ranking of Department distribution.

**Distribution of Papers by Types of Documents**

**Table 7:** Distribution of Publications by Document Type

| S/N | Document Type | Records | % | TLCS | TGCS |
|---|---|---|---|---|---|
| 1 | Article | 5249 | 82.62 | 3,813 | 73,748 |
| 2 | Review | 458 | 7.21 | 272 | 10,369 |
| 3 | Meeting Abstract | 295 | 4.64 | 2 | 56 |
| 4 | Editorial Material | 120 | 1.89 | 81 | 1,119 |
| 5 | Article; Early Access | 102 | 1.61 | 0 | 167 |
| 6 | Correction | 43 | 0.68 | 0 | 4 |
| 7 | Letter | 24 | 0.38 | 9 | 131 |
| 8 | Article; Proceedings Paper | 15 | 0.24 | 14 | 464 |
| 9 | Book Review | 13 | 0.20 | 0 | 1 |
| 10 | Article; Data Paper | 7 | 0.11 | 0 | 60 |
| 11 | Review: Early Access | 6 | 0.09 | 0 | 23 |
| 12 | Art Exhibit Review | 4 | 0.06 | 1 | 2 |
| 13 | Article: Retracted Publication | 4 | 0.06 | 0 | 30 |
| 14 | Correction: Early Access | 2 | 0.03 | 0 | 2 |
| 15 | Editorial Material; Early Access | 2 | 0.03 | 0 | 0 |
| 16 | Film Review | 2 | 0.03 | 0 | 0 |
| 17 | News Item | 2 | 0.03 | 1 | 25 |
| 18 | Retraction | 2 | 0.03 | 0 | 0 |
| 19 | Biographical-Item | 1 | 0.02 | 0 | 0 |
| 20 | Meeting | 1 | 0.02 | 0 | 0 |
| 21 | Reprint | 1 | 0.02 | 0 | 1 |
| **Total** | | **6,353** | **100** | **4,193** | **86,202** |

TLCS = Total Local Citation Score; TGCS = Total Global Citation Score

Table 7 describes the document-type contribution of research from the University of Nigeria. Based on the analysis, it was discovered that 21 document-type research contributions were made. The significant document type appears in the form of Articles, at (5,249), followed by Review (458), Article; Proceedings Paper (15), Meeting Abstract (259), while the minor document type records at (0.02%), which comprise Biographical-Item, meeting; Reprint. This record reveals that the article has the highest document type contribution of research at (82.62%) followed by Review, which scored (7.21%), and Article Proceedings Paper at (4.64%).

### Most Relevant Countries by Corresponding Authors

**Table 8:** Geographic Distribution of Corresponding Authors and International Research Engagement

| S/N | Country | Articles | Articles % | SCP | MCP | MCP % |
|---|---|---|---|---|---|---|
| 1 | Nigeria | 4,445 | 70 | 3,161 | 1,284 | 28.9 |
| 2 | USA | 266 | 4.2 | 0 | 266 | 100 |
| 3 | China | 244 | 3.8 | 0 | 244 | 100 |
| 4 | United Kingdom | 219 | 3.4 | 0 | 219 | 100 |
| 5 | South Africa | 197 | 3.1 | 1 | 196 | 99.5 |
| 6 | Germany | 74 | 1.2 | 1 | 73 | 98.6 |
| 7 | Canada | 66 | 1 | 0 | 66 | 100 |
| 8 | Australia | 54 | 0.8 | 0 | 54 | 100 |
| 9 | India | 51 | 0.8 | 0 | 51 | 100 |
| 10 | Japan | 41 | 0.6 | 0 | 41 | 100 |
| 11 | Italy | 33 | 0.5 | 0 | 33 | 100 |
| 12 | Malaysia | 28 | 0.4 | 0 | 28 | 100 |
| 13 | Poland | 26 | 0.4 | 0 | 26 | 100 |
| 14 | Iran | 23 | 0.4 | 0 | 23 | 100 |
| 15 | Cameroon | 22 | 0.3 | 0 | 22 | 100 |
| 16 | Netherlands | 19 | 0.3 | 0 | 19 | 100 |
| 17 | Benin | 18 | 0.3 | 13 | 5 | 27.8 |
| 18 | Ethiopia | 17 | 0.3 | 0 | 17 | 100 |
| 19 | Gambia | 13 | 0.2 | 0 | 13 | 100 |
| 20 | New Zealand | 12 | 0.2 | 0 | 12 | 100 |

SCP=Single Country Publications; MCP=Multiple Country Publications

Table 8 identifies the nations with the most publications and corresponding authors. The SCP and MCP metrics measure the number of single corresponding authors and multiple corresponding authors. The frequency and MCP ratio provide light on the contributions made by writers from each nation to the field. The list demonstrates that the two nations with the most significant number of corresponding authors in the area of big data are Nigeria and the United States. The United States has 266 articles, whereas Nigeria has 1284 papers with corresponding authors. This shows that considerable research is being pioneered in these two nations.

Furthermore, both countries have a higher-than-average MCP ratio, which suggests that authors from both nations are more likely to have numerous corresponding authorships. The list highlights the fact that extensive data research is multidisciplinary. The nations featured in the list come from several continents; for example, matching writers may be found for Australia, India, Japan, Italy, and South Africa. This demonstrates the broad spectrum of big data applications and uses, from medical applications to intelligent transportation. In conclusion, this list of nations and their corresponding authors sheds light on the worldwide significance of significant data research. It also demonstrates the multidisciplinary nature of ample data research and the need for international collaboration among scholars to investigate its possible uses.

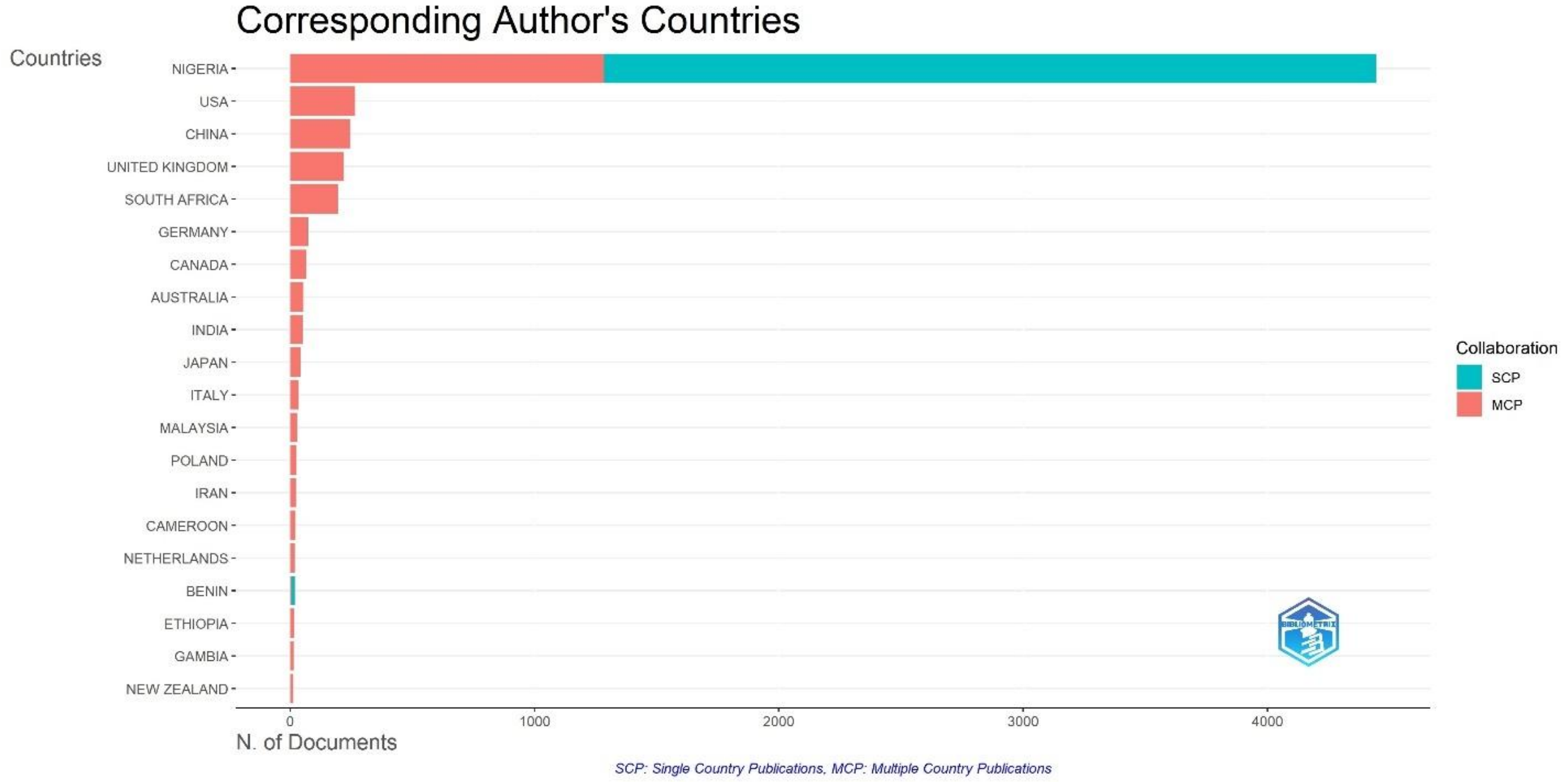

Figure 4: Corresponding Author's Countries

## DISCUSSION

The scientometric analysis of the University of Nigeria's research output from 2014 to 2023 shows a steady yet uneven growth pattern, with 6,353 publications indexed in the Web of Science database. The rise in output after 2018 reflects an expanding research culture and increased institutional participation in global scholarship. Similar growth trends have been reported in institutional studies by Ahmad and Batcha (2019) and Darmadji et al. (2018), where collaboration and institutional support contributed to increased productivity. However, fluctuations in annual output and the 8.6% annual growth rate suggest challenges related to funding and infrastructural constraints, which are common in African higher education systems (Nweze, 2021).

The dominance of prolific authors such as Mokdad, Fischer, and Ezema (Year(s)) has significantly strengthened research visibility, particularly through large collaborative studies. The average of 12.5 co-authors per publication indicates a strong collaborative culture consistent with findings by Mahala and Singh (2021), who observed that multi-authored publications tend to achieve higher citation impact.

Publishing in prominent journals such as The Lancet, Medicine, and Applied Water Science demonstrates engagement with high-impact outlets and contributes to institutional visibility. Similar observations were reported by Ahmad and Ubi (2025), where publication in reputable journals enhanced global recognition. Although the university accumulated 86,202 citations with an average of 13.57 citations per paper, comparison with institutions such as the University of Ibadan (Ahmad & Nkatv, 2025) suggests opportunities to further improve citation influence through strategic publication planning.

Collaboration patterns reveal strong partnerships with Nigerian institutions, including Nnamdi Azikiwe University and the University of Ibadan, alongside international collaborators such as Cairo University and the London School of Hygiene & Tropical Medicine. The international co-authorship rate of 46.88% aligns with Mahala and Singh (2021), emphasizing collaboration as a driver of research visibility. However, limited intra-African collaboration highlights an area for future improvement.

Departmental productivity is dominated by medical and science-related units, particularly the College of Medicine and the University of Nigeria Teaching Hospital, supported by departments such as Pure and Industrial Chemistry and Physics. Similar disciplinary concentration has been noted in institutional bibliometric studies (Das et al., 2021; Ahmad & Ubi, 2025), often linked to stronger funding opportunities in health and applied sciences.

Research articles constitute the majority of document types, confirming a preference for empirical scholarly communication. Comparable findings were reported by Kumar and Dora (2012) and Das et al. (2021), identifying journal articles as the primary medium of academic dissemination. The distribution of corresponding authors further shows Nigeria leading, followed by the United States and other developed research systems, reflecting expanding global engagement similar to findings by Ahmad and Nkatv (2025).

Overall, the University of Nigeria has demonstrated considerable progress in productivity, collaboration, and citation impact. Sustained investment in research funding, mentorship, and open-access publishing, alongside strengthened intra-African partnerships and improved research infrastructure, will further enhance its global research standing.

## CONCLUSION AND RECOMMENDATIONS

### Conclusion

This scientometric assessment of the University of Nigeria's research productivity from 2014 to 2023 provides a clear and evidence-based understanding of the institution's scholarly trajectory across the decade. The analysis shows that research output grew steadily but with fluctuations, culminating in 6,353 publications, with a notable surge after 2018 that reflects an improving research culture and increasing engagement in global scholarship. In line with the second objective, the evaluation of authorship revealed that scholars such as Mokdad A.H., Fischer F., and Ezema F.I. emerged as the most prolific and influential contributors, supported by a robust culture of multi-authorship that averaged 12.5 authors per article. This underscores strong collaborative practices within the university.

Regarding the third objective, the study identified high-impact journals like *The Lancet*, *Medicine*, and *Applied Water Science* as major outlets for UNN researchers, demonstrating notable visibility in reputable publication sources. Analysis of collaborative networks further showed that the university maintained strong institutional partnerships with Nnamdi Azikiwe University, University of Ibadan, Cairo University, and the London School of Hygiene & Tropical Medicine, aligning with the fourth objective by highlighting both national and international collaboration patterns.

In relation to departmental contributions, the College of Medicine, the University of Nigeria Teaching Hospital, and departments such as Pure and Industrial Chemistry, Physics, and Psychology led in publication output, demonstrating that medical and science-related units are the backbone of UNN's research strength, directly addressing the fifth objective. The examination of document types found that research articles overwhelmingly dominated with 82.62%, confirming a strong emphasis on empirical research outputs, thereby satisfying the sixth objective. Finally, the analysis of corresponding author countries showed that Nigeria and the United States contributed the largest share, with a 46.88% international co-authorship rate, reflecting a broadening global research footprint consistent with the seventh objective.

Collectively, the findings highlight UNN's growing research productivity, expanding collaborative reach, and increasing citation impact. Despite periodic fluctuations in annual output, the decade-long performance demonstrates resilience, rising global relevance, and significant contributions across multiple disciplines.

## Recommendations

To strengthen future research performance, the University of Nigeria should prioritize consistent research funding, expand mentorship programs for early-career scholars, and encourage greater participation in open-access publishing to improve visibility. Strengthening intra-African collaboration, upgrading research infrastructure, and supporting interdisciplinary projects will help stabilize annual output patterns and deepen the university's global engagement. Enhancing departmental capacity, especially in non-science fields, and promoting international mobility programs will further elevate its scholarly impact and ensure sustained growth across all indicators.